\documentclass[12pt]{iopart}
  \usepackage[dvips]{color}
  \usepackage{newlfont}
  \usepackage{graphicx}
  \usepackage{amssymb}
  \usepackage[latin1]{inputenc}
  \usepackage{float} 
\begin{document}

\title{Low-resolution measurements induced classicality}
\author{R M Angelo}

\address{Departamento de Física, Universidade Federal do Paraná, 81531-990, Curitiba, PR, Brazil}
\ead{renato@fisica.ufpr.br}

\begin{abstract}
The classical limit of quantum mechanics is discussed for closed quantum systems in terms of observational aspects. Initially, the failure of the limit $\hbar\to 0$ is explicitly demonstrated in a model of two quantum mechanically interacting oscillators by showing that neither quantum expectations reduce to Newtonian trajectories nor entanglement vanishes. This result suggests that the quantum-to-classical transition occurs only at an approximative level, which is regulated by the low accuracy of the measurements. In order to verify the consistence of these ideas we take into account the experimental resolution of physical measurements by introducing a discretized formulation for the quantum structure of wave functions. As a result, in the low-resolution limit the quasi-determinism is recovered and hence the quantum-to-classical transition is shown to occur adequately. Other puzzling problems, such as the classical limit of quantum superpositions and nonlocal correlations, are naturally address as well.
\end{abstract}

\maketitle

\section{Introduction}
Understanding the emergence of classical behavior from the quantum substratum is probably one of the oldest challenges of the quantum theory. While it is a well known fact that classical physics dramatically fails in explaining the microscopic world, we cannot say that there exists a decisive proof attesting the
universality of quantum mechanics as a theory capable of accounting for all aspects of the classical world. The limit $\hbar\to 0$ and the Ehrenfest theorem~\cite{ehrenfest} have recurrently been proved not to be sufficient to guarantee the classical limit both mathematically and conceptually
\cite{berry89,berry95,ballentine94,adelcio03,ballentine04,angelo05,angelo07}. More modern approaches such as the environment induced decoherence (EID) program also have been claimed to present some conceptual difficulties (see~\cite{angelo07,wiebe05} and references therein for more detailed discussions), as for instance: i-) The incapability of diffusive EID in restraining the wave function spreading and hence recovering the classical determinism and ii-) the apparently paradoxical presence of entanglement -- an exclusively quantum resource -- in semiclassical regimes.

Recently, approaches based on observational aspects have been considered in order to address the quantum-to-classical transition in closed spin systems~\cite{kofler07} as well as in bosonic systems under diffusive decoherence~\cite{angelo07}. As has been suggested by these works, the apparent Newtonian motion of a falling ball in the vicinity of the Earth is nothing more than an effect of both the low resolution power of our experimental apparatus and the short time scales usually involved in the phenomenon. More powerful devices would reveal the dispersion associated with the wave function spreading in a run of similarly prepared experiments, thus contradicting our classical beliefs on the Newtonian determinism. On the other hand, in experiments of the macroscopic world, when we are interested, e.g., in describing the motion of the Moon, arbitrarily high resolutions is not needed anymore. In this case, another scale is required which will inevitably ignore the fine quantum details of the macroscopic body dynamics. In this case, a kind of deterministic behavior (``quasi-deterministic"~\cite{angelo07}, to be precise) is observed to exist. According to this view, classical behavior only exists as an approximated notion derived from low-resolution measurements.

This work is mainly devoted to demonstrate the relevance of observational aspects in addressing the quantum-to-classical transition. We start by discussing the inefficiency of the formal limit $\hbar\to 0$ in describing the emergence of the classical structure from the quantum formalism in an analytical model of two oscillators coupled via a nonlinear phase interaction. The existence of nonvanishing entanglement in semiclassical regime is analytically verified and its role in enhancing subsystem classicality is pointed out. Experimental aspects are shown to be essential in explaining the quantum-to-classical transition and an approach is proposed which allows us to take into account the experimental resolution directly in the formal structure of quantum mechanics. As a result, in the limit of low resolutions classical physics is shown to arise.

\section{Failure of the $\hbar\to 0$ limit}
In this section, the insufficiency of the mathematical limit $\hbar\to 0$ is explicitly demonstrated for a model of two anharmonic oscillators with a nonlinear coupling. This model has been used to describe phase interactions between modes of bosonic fields in the presence of nonlinear Kerr media
\cite{semiao05} and BECs species~\cite{sanz03}. The Hamiltonian reads
\begin{eqnarray}\label{Hq}
\hat{H}=\omega_1 \hat{h}_1 + g_1\hat{h}_1^2 + \omega_2\hat{h}_2+g_2\hat{h}_2^2+g\hat{h}_1\hat{h}_2,
\end{eqnarray}
where $\hat{h}_{k}\equiv\hbar\left(\hat{n}_{k}+\frac{1}{2} \right)$. The numbers $\omega_{k}$ and $g_k$ denote, respectively, the normal frequency and the nonlinearity parameter of the $k$-th mode and $g$ is the phase coupling parameter. In order to give a best chance to the limit $\hbar\to 0$ let us consider disentagled coherent states at $t=0$, i.e., $|\psi(0)\rangle=|z_{10}\rangle\otimes|z_{20}\rangle$, with the usual parametrization $z_{k0}=(q_{k0}+\imath p_{k0})/\sqrt{2\hbar}$. The classical Hamiltonian, which produces (\ref{Hq}) through the method of ordered symmetric quantization
\cite{angelo03}, reads
\begin{eqnarray}\label{Hcl}
\mathcal{H}=\omega_1 h_1 + g_1 h_1^2 + \omega_2 h_2+g_2 h_2^2+g h_1 h_2,
\end{eqnarray}
where $h_k\equiv\frac{1}{2}\left(p_k^2+q_k^2\right)$. The canonical pair $(q_k,p_k)$, with dimensional unit of angular momentum, relates to the usual canonical pair position-momentum $(Q_k,P_k)$ through the canonical transformation $(q_k,p_k)=(Q_k\sqrt{m_k\omega_k},P_k/\sqrt{m_k\omega_k})$.
Classical solutions are readily obtained by noticing that $h_1$ and $h_2$ are constants of motion. The result is written in matrix form as follows:
\begin{eqnarray}\label{Rcl}
&R_{k}^{cl}(t)=M[\phi_{k}^{cl}(t)]R_{k0},&\\
&\phi_{k}^{cl}(t)=\omega_k t+g_k S_k t+\frac{g S_j t}{2},& \qquad (j\neq k)
\end{eqnarray}
where $S_{k}\equiv (p_{k0}^2+q_{k0}^2)$, $R_{k}^{cl}(t)\equiv\left({q_{k}(t)\atop p_{k}(t)}\right)$, $R_{k0}=\left({q_{k0}\atop p_{k0}}\right)$, and $M$ is the usual $2\times 2$ rotation matrix.

In order to obtain the quantum solutions we firstly write the time evolved density operator in the Fock basis:
\begin{eqnarray}\label{rho}
\hat{\rho}(t)&=&e^{-|z_{10}|^2- |z_{20}|^2}\sum\limits_{n, n' \atop m, m'}
\frac{z_{1t}^n (z_{1t}^*)^{n'}z_{2t}^m(z_{2t}^*)^{m'}}{\sqrt{n!\,n'!\,m!\,m'!}}
e^{-\imath \hbar g t (nm-n'm')} \nonumber \\
&& \times \,e^{-\imath \hbar g_1 t (n^2-n'^2)}e^{-\imath \hbar g_2 t (m^2-m'^2)}
|n\rangle\langle n'|\otimes|m\rangle\langle m'|,
\end{eqnarray}
where $z_{kt}=z_{k0}e^{-\imath \Omega_k t}$, with $\Omega_k\equiv \omega_k+\hbar g_k+\hbar g/2$. Using this result one may show that the expectation values of position
and momentum can be written as
\begin{eqnarray} \label{Rq}
&\langle \hat{R}_{k}\rangle (t)=A_k(t)M[\phi_k(t)]R_{k0},&\\
&\phi_{k}(t)=\omega_k t+2\hbar g_k t+\frac{\hbar g t}{2}+\frac{S_k}{2\hbar}\sin\left(\hbar g_k
t\right)+\frac{S_j}{2\hbar}\sin\left(\hbar g t\right),&  \\
&A_k(t)=\exp\left[-\frac{2S_k}{\hbar}\sin^2\left(\frac{\hbar g_k t}{2}\right)
-\frac{2S_j}{\hbar}\sin^2\left(\frac{\hbar g t}{2}\right)\right],&
\end{eqnarray}
where $\langle \hat{R}_k\rangle \equiv\left(\langle \hat{q}_k \rangle \atop \langle \hat{p}_k\rangle
\right)$.

Now let us consider the following strict mathematical limit:
\begin{eqnarray}
\lim\limits_{\hbar\to 0}\langle \hat{R}_{k}\rangle(t)=M[\phi_k^{cl}(t)-g_kS_kt/2]R_0=M[-g_kS_k
t/2]R_{k}^{cl}(t).
\end{eqnarray}
The failure of this limit in reproducing the classical trajectory (\ref{Rcl}) occurs because
nonlinearities (present whenever $g_k\neq 0$) induce quantum superpositions in the evolved quantum state, as we shall see next. Of course, this is not an appropriate {\em physical limit} since neither $\hbar$ can assume arbitrarily small values (it actually is a constant of nature) nor the term $\hbar\, t$ in equation~(\ref{Rq}) is always small. Nevertheless, this is an explicit demonstration of the incommutability between the limits $\hbar\to 0$ and $t\to\infty$ as has been claimed by Berry~\cite{berry89,berry95}.

Difficulties with this limit also appear in the entanglement dynamics, which can be monitored in closed pure bipartite systems with  the {\em linear entropy}, defined as $\mathcal{E}(t)=1-\textrm{Tr}_1\hat{\rho}_1^2(t)$, where
$\hat{\rho}_1(t)=\textrm{Tr}_2|\psi(t)\rangle\langle \psi(t)|$. The result for the situation under
consideration is
\begin{eqnarray}
\mathcal{E}(t)=1-e^{-2|z_{10}|^2}\sum\limits_{n=0}^{\infty} \sum\limits_{n'=0}^{\infty}\frac{|z_{10}|^{2n}}{n!} \frac{|z_{20}|^{2n'}}{n'!}e^{-4|z_{20}|^2\sin^{2}\left[\frac{\hbar g t}{2}(n-n') \right]}.
\end{eqnarray}
Within an arbitrarily short time scale one can expand the exponential and proceed with some
algebraic manipulation in order to obtain
\begin{eqnarray}\label{Eapprox}
\mathcal{E}(t)\simeq 2\left(|z_{10}|\,|z_{20}|\,\hbar\, g\, t  \right)^2=(S_1\, g\, t)(S_2\,g\,t).
\end{eqnarray}
Clearly, $\lim_{\hbar\to 0}\mathcal{E}\neq 0$ and hence the classical limit for the entanglement, which is supposed to be $\mathcal{E}=0$, is not correctly predicted by this approach, in agreement with the results obtained in~\cite{angelo05}. These results -- rigorously obtained in an analytical problem -- show that classical maths cannot be expected to emerge from the quantum structure in some formal limit. In addition, in the light of the results reported in~\cite{angelo07}, one may assert that decoherence cannot make the situation better for conservative systems: Although it can destruct the entanglement between the subsystems it is not capable of restraining the wave function spreading, thus not recovering Newtonian trajectories and quasi-determinism.

The above discussion obligates us to conceive that the classical behavior arises from quantum substratum only as an approximation. Specifically for the system studied here this approximative
character is qualified by the conditions
\begin{eqnarray}\label{conditions}
\hbar\,g_k\,t\ll \frac{\hbar}{S_k}\ll 1\qquad \textrm{and}\qquad \hbar\,g\,t\ll\frac{\hbar}{S_k} \ll 1,
\end{eqnarray}
the former (later) accounting for nonlinearity (entanglement) effects. Under these conditions, which clearly point out the need for short time scales, straightforward calculations show that we can approximate: $\langle \hat{R}_k\rangle (t)\approx R_k^{cl}(t)$ and $\mathcal{E}(t)\approx 0$.

Once the approximative character of the quantum-to-classical transition has been demonstrated, the relevance of the experimental aspects in qualifying the classical limit has properly been put in evidence. This motivates the observational approach we shall present in this paper. Before continuing, however, we address the question concerning the unavoidable presence of entanglement even in large actions regimes.

\subsection{The role of the entanglement for the classical limit}

It has been shown that apart from the trivial situation in which $t=0$ entanglement is always formally present between the subspaces. In this sense one may wonder whether it is indeed possible to recover the classical limit from the quantum substratum. On the other hand, we can suspect -- based on the well known fact that decoherence is fundamentally mediated by entanglement -- that entanglement may play some role in the classical limit of closed systems with few degrees of freedom. Now we show that this is indeed the case.

Let us consider the reduced density matrix $\hat{\rho}_1(t)=\textrm{Tr}_2|\psi(t)\rangle\langle\psi(t)|$.
From (\ref{rho}) we obtain
\begin{eqnarray}\label{rho1}
\hat{\rho}_1(t)=e^{-|z_{10}|^2}\sum\limits_{n,n'}\frac{z_{1t}^n(z_{1t}^*)^{n'}}{\sqrt{n!\,n'!}}
\mathcal{D}_{nn'}(t) e^{-\imath \hbar g_1 t(n^2-n'^2)} |n\rangle\langle n'|,
\end{eqnarray}
where
\begin{eqnarray}\label{D}
\mathcal{D}_{nn'}(t) \equiv e^{-\frac{S_2}{\hbar}\sin^2\left[\frac{\hbar g t}{2} (n-n')\right]}\,e^{-\imath\frac{S_2}{\hbar}\sin\left[\hbar\,g\,t\,(n-n')\right]},
\end{eqnarray}
with $S_2=(p_{20}^2+q_{20}^2)$. Consider for a moment that $g=0$ (so that $\mathcal{D}_{nn'}=1$). In this case, there is no interaction and hence no entanglement dynamics. Following~\cite{banerji01}, we now consider instants
$t_{r,s}\equiv \frac{r}{s} \frac{\pi}{g_1\hbar}$, $r$ and $s$ being relatively prime numbers, in terms of
which we can write a discrete Fourier transform as follows:
\begin{eqnarray}\label{transf}
e^{-\imath \pi n^2 \frac{r}{s}}=\sum\limits_{q=0}^{l-1}a_q^{(r,s)}e^{-\imath 2 \pi n \frac{q}{l}},
\end{eqnarray}
where
\begin{eqnarray}
a_q^{(r,s)}=\frac{1}{l}\sum\limits_{k=0}^{l-1}e^{-\imath \pi k \left(k\frac{r}{s}-2\frac{q}{l}\right)}
\end{eqnarray}
and
\begin{eqnarray}
l=\left\{
\begin{array}{l}
\textrm{$s$, for $r$ odd and $s$ even (or vice-versa)} \\ \textrm{$2s$, for $r$ and $s$ odd integers.}
\end{array} \right.
\end{eqnarray}
Using these relations we may rewrite (\ref{rho1}) as
\begin{eqnarray}
\hat{\rho}_1(t_{r,s})=|\Psi_{r,s}\rangle\langle\Psi_{r,s}|
\end{eqnarray}
where
\begin{eqnarray}
|\Psi_{r,s}\rangle = \sum\limits_{q=0}^{l-1}a_{q}^{(r,s)}|z_{1t_{r,s}}e^{-\imath 2\pi\frac{q}{l}}\rangle.
\end{eqnarray}
State $|\Psi_{r,s}\rangle$ is a kind of generalized cat state: a symmetrical quantum superposition of several coherent states centered at points a distance $|z_{1t_{r,s}}|$ from the origin of the complex plane, with an angular separation $2\pi/l$ between neighbors. Then, in the absence of entanglement the reduced density matrix evolves (retaining its initial purity) to a manifest nonclassical quantum superposition.

The situation is rather different when the entanglement dynamics is switched on ($g\neq 0$). In this case $\mathcal{D}_{nn'}(t_{r,s})$ yields an attenuation in the off--diagonal terms of $\hat{\rho}_1$, thus destructing its purity. As a consequence, the coherence of the reduced density operator is partially destructed and hence the formation of the generalized cat state is prevented. More importantly, in (\ref{rho1}) we can clearly observe the role played by the large actions limit in this scenario: If $\hbar/S_2\ll 1$,
then $\mathcal{D}_{nn'}(t)\approx \delta_{n,n'}$ even for arbitrarily large $t$ and hence
\begin{eqnarray}\label{stat}
\hat{\rho}_1(t)\approx \sum_n e^{-|z_{10}|^2}\frac{|z_{10}|^{2n}}{n!}|n\rangle\langle n|,
\end{eqnarray}
which manifestly describes a statiscal mixture. In words, the classical limit of oscillator 2 implies, via entanglement, the {\em statistical classical limit} of oscillator 1 in the sense of the Liouville theory. That is, entanglement -- the quantum resource associated with {\em nonlocal} correlations -- ironically is the responsible for ensuring the nonexistence of superposition states for subsystems.

Before closing this section, some comments are in order. Firstly, our analysis, which is focused on a bipartite system with only a few degrees of freedom, shows an example in which decoherence arises from the classicality related to one of the subsystem ($\hbar/S_2\ll 1$) and not from the coupling with the many degrees of freedom of an external thermal environment, as is usually claimed. Secondly, the existence of entanglement in classical regime seems to have been well justified: It prevents the cat state formation and guarantees the classical notion of statistics (Liouvillian classical limit) during the entire evolution of the system. Thirdly, in spite of the simplicity of our model we believe it is able to capture the essential features of nondissipative decoherence that is expected to take place in the dynamics of most open conservative systems (see~\cite{angelo07,semiao05} for more detailed discussions). Finally, as the Newtonian classical limit is concerned the conditions for the quasi-determinism emergence~\cite{angelo07} turn out to be those given by equation~(\ref{conditions}). In this case, the quantum-to-classical transition becomes dependent on our incapability of detecting the wave function spreading. This conducts us to the next main point assessed in this paper.

\section{Classicality emerging from observational limitations}

The role played by the experimental resolution in the diagnostic of the quantum-to-classical transition has recently been suggested in different contexts~\cite{angelo07,kofler07}. Here we provide general demonstrations of how observational aspects can physically explain the emergence of classicality from the quantum substratum. The best of our knowledge this is the first time such a task is carried out.

\begin{figure}[ht]
\centerline{\includegraphics[scale=0.35]{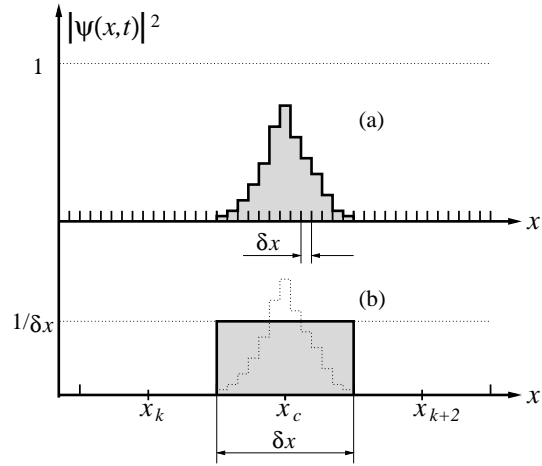}}
\caption{(a) In the regime of {\em high resolutions} ($\delta x\ll \Delta x$) the wave function width is experimentally accessible and the continuum wave function appears. (b) On the other hand, for {\em low resolutions} ($\delta x\gg \Delta x$) the apparatus cannot resolve the quantum dispersion and the statistics resembles a normalized Kronecker delta function.}
\label{fig1}
\end{figure}

Our strategy consists in inserting the experimental information directly in the mathematical structure of the quantum formalism and then show that this indeed produces the expected classical results in some limit. In order to understand how this can be accomplished we focus on the situation depicted in figure~\ref{fig1}, where the square modulus of a one-particle wave function is given, for instance, for a wave-matter diffraction experiment. We consider that the screen (in the $x$ axis) is completely covered by rectangular detectors, each one with width $\delta x$. Whenever a particle reaches a point of the $k$-th detector a click is registered and the position $x_k$ is attributed to the measurement. This scheme yields a discretization of the observed statistics. As a consequence, from an observational viewpoint, it
is possible to think of positions as discrete numbers, $x_k$, conveniently located at the center of the respective detector $k$, so that $x_{k+1}-x_k=\delta x$. According to figure~\ref{fig1}
the normalization condition can be expressed as
\begin{eqnarray}
\sum\limits_{k=-\infty}^{\infty}\delta x\,|\psi(x_k,t)|^2=1,
\end{eqnarray}
for any value of $\delta x$. Note that in our approach the experimental resolution is readily identified with the detector width $\delta x$. For small enough detectors $\delta x$ becomes a differential displacement and the usual normalization condition is recovered. In this case, the wave function width (given by the variance $\Delta x(t)=\left[\langle\hat{x}^2 \rangle-\langle \hat{x}\rangle^2\right]^{1/2}$) may be resolved and, as a consequence, the quantum character of the dynamics at that instant $t$ is mostly identified (see figure~\ref{fig1}-a). Realistic detectors, however, cannot be thought of as arbitrarily small devices. In fact, according to quantum measurement theory (see, e.g.,~\cite{walls}) they have to be modeled as macroscopic (classical in some sense) objects. Then, let us consider a short time scale within which an initially narrow wave function remains localized such that $\Delta x(t)< \delta x$. Now the wave function dispersion cannot be resolved and the Newtonian mechanics suffices to predict the position of the particle at the instant $t$ (see figure~\ref{fig1}-b). This analysis shows in which sense {\em low resolution} implies {\em classical quasi-deterministic behavior} and allows us to
introduce the classical limit in terms of the prescription
\begin{eqnarray}\label{psiCL}
\delta x
\,\,\psi^{*}(x_i,t)\psi(x_j,t)\to \delta_{x_i,x_c}\delta_{x_j,x_c},
\end{eqnarray}
where $\delta_{x_{i(j)},x_c}$ denotes the Kronecker delta function and $x_c$ stands for the position at which the statistics is centered. Note that the singular character of the classical limit~\cite{berry89,berry95} is introduced here through experimental justifications, here derived from the scheme proposed in figure~\ref{fig1}, instead of strictly mathematical requests such as
$\hbar/S=0$.

The next step towards the construction of our discretized resolution-based approach for the
quantum formalism is to rewrite the scalar product between eigenstates of the position operator as
\begin{eqnarray}\label{xixj}
\langle x_i|x_j\rangle=\frac{\delta_{x_i,x_j}}{\delta x},
\end{eqnarray}
which suitably tends to the Dirac delta function in the high resolution limit ($\delta x\to 0$)
and possesses the correct dimensional unit, as required by the completness relation
\begin{eqnarray}\label{1}
\sum_{k=-\infty}^{\infty}\delta x\,|x_k\rangle\langle x_k|=\mathbf{1}.
\end{eqnarray}
The momentum in the position representation will be given by
\begin{eqnarray}\label{px}
\langle x_k|\hat{p}|\psi\rangle=\frac{\hbar}{\imath}\frac{\delta \psi(x_k)}{\delta x},
\end{eqnarray}
where $\delta f(x_k)=f(x_k+\delta x)-f(x_k)=f(x_{k+1})-f(x_k)$. Now, using (\ref{1}) and
(\ref{px}) one may show straightforwardly that
\begin{eqnarray}
\langle\psi|\hat{x}\hat{p}|\psi\rangle-\langle\psi|\hat{p}\hat{x}|\psi\rangle=\imath\hbar\sum_{k}\delta
x\,\,\psi^{*}(x_k)\,\psi(x_k+\delta x).
\end{eqnarray}
The adequacy of our approach in describing the quantum-to-classical transition is attested by this expression. In the limit of high resolutions (quantum regime), i.e., for $\delta x\to 0$, the usual formula $\langle\psi| [\hat{x},\hat{p}]|\psi\rangle=\imath\hbar$ is recovered in virtue of the normalization condition $\int dx \,\psi^*(x)\psi(x)=1$. However, in the low resolution regime (classical regime) by (\ref{psiCL}) we see that $\delta x\,\,\psi^*(x_k)\psi(x_{k+1})=0$ and hence that $\langle\psi| [\hat{x},\hat{p}]|\psi\rangle=0$. Between these extrema, i.e., for arbitrary resolutions, we may write $\psi(x_k+\delta x)=\phi(x_k)$ and use the Schwartz inequality, $|\langle\psi|\phi\rangle|^2\le
\langle\psi|\psi\rangle\langle\phi|\phi\rangle=1$, in the position representation to show that
\begin{eqnarray}
\left|\sum_{k}\delta x\,\,\psi^{*}(x_k)\,\psi(x_k+\delta x)\right|\le 1,
\end{eqnarray}
the equality holding for $\delta x\to 0$. These results allow one to write
\begin{eqnarray}
\textrm{(classical)}\qquad 0\le \left|\frac{\langle\psi
|[\hat{x},\hat{p}]|\psi\rangle}{\imath\hbar}\right|\le 1 \qquad \textrm{(quantum)}.
\end{eqnarray}
This theoretical result shows how the experimental resolution can alter the predictions causing them to range from a strictly classical to a quantum regime, passing through a semiclassical one. Next we show how such a experimental resolution based formalism can address several puzzling aspects of the quantum-to-classical transition.

\subsection{Low-resolution limit of expectation values}
Now we focus on demonstrating how Newtonian mechanics emerges from the Heisenberg equations in the limit of low resolutions. For a conservative one-particle system described by
$\hat{H}=\hat{p}^2/2m+V(\hat{x})$ Ehrenfest showed~\cite{ehrenfest} that we may write, without any approximation, that $m \,d^2
\langle\hat{x}\rangle/dt^2=-\langle \partial_{\hat{x}}V(\hat{x})\rangle$.
In our approach this is written as
\begin{eqnarray}
m\frac{d^2}{dt^2}\sum_k\delta x\, |\psi(x_k,t)|^2\, x_k=-\sum_k\delta
x\, |\psi(x_k,t)|^2\,\frac{\delta V(x_k)}{\delta x}.
\end{eqnarray}
Using (\ref{psiCL}) for the low-resolution limit we find
\begin{eqnarray}
m \frac{d^2x_c}{dt^2}=-\frac{\delta V(x_c)}{\delta x},
\end{eqnarray}
which is our resolution-based formulation of Newton's second law. Even though this result could be regarded in some sense as an alternative demonstration of the Ehrenfest theorem it actually is an indication of the relevance of observational aspects in addressing the quantum-to-classical transition.

It is worth emphasizing that we do not reformulate in any aspect the dynamics of the quantum theory. Only the kinematic structure associated with physical projections has been adapted to fit the conceptual framework associated with the experimental aspects of physics. Also, for
the sake of simplicity, we have chosen to work with the one dimensional configuration space. The generalization of these ideas for a phase space formalism, such as the Weyl-Wigner formulation, will be carried out elsewhere.

\subsection{Low-resolution limit of quantum superpositions}
\begin{figure}[htb]
\centerline{\includegraphics[scale=0.7]{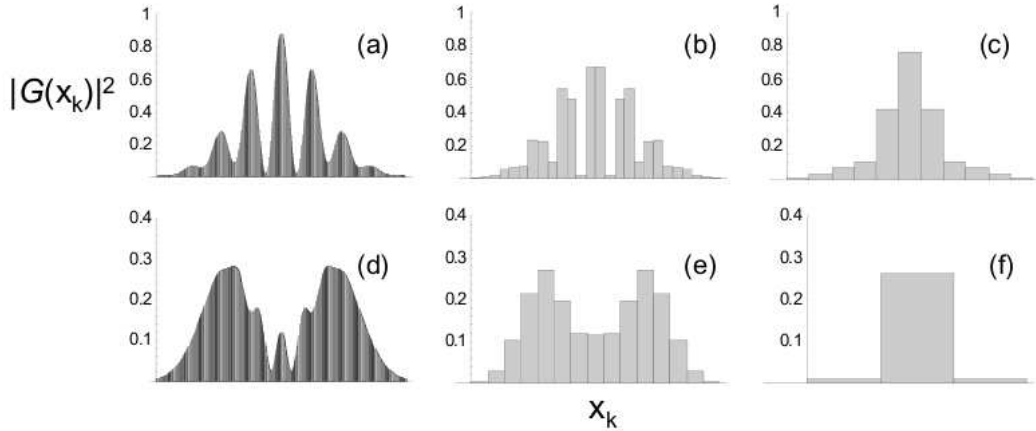}}
\caption{Density probability $|G(x_k)|^2$ as a function of $x_k$ for $\lambda=0.2\,b$ and $b=1$ [a.u.]. Discretized positions are written as integer multiples of the experimental resolution, i.e., $x_k=k\,\delta x$. In the first line, $q=b$ (near peaks) and (a) $\delta x= 0.01\,b$, (b) $\delta x=0.2\,b$, and (c) $\delta x=0.5\,b$ whereas in the second, $q=3\,b$ (far apart peaks) and (d) $\delta x= 0.01\,b$, (e) $\delta x=0.5\,b$, and (f) $\delta x=3.5\,b$.}
\label{fig2}
\end{figure}

Now we investigate the applicability of our approach in describing the classical limit of a paradigmatic behavior of the quantum world, namely, a quantum superposition
state. Let us consider a superposition of coherent states given by
\begin{eqnarray}
G(x)=\frac{1}{\cal{N}}\big[\varphi_+(x)+\varphi_-(x)\big],
\end{eqnarray}
where $\cal{N}=\left[2(1+e^{-2|z|^2})\right]^{1/2}$, $\varphi_{\pm}(x)=\langle x|\pm z\rangle$, and
\begin{eqnarray}
\langle x| z\rangle =
\frac{1}{\sqrt{b\sqrt{\pi}}}\exp\left[-\frac{(x-q/2)^2}{2b^2}\right]
e^{\frac{\imath(x-q/2)}{\lambda}},
\end{eqnarray}
with $z=\frac{1}{\sqrt{2}}(\frac{q}{b} + \imath \frac{b}{\lambda})$, $b=\sqrt{\hbar/m\omega}$, and
$\lambda\equiv \hbar/p$. Figure~\ref{fig2} illustrates the emergence of classicality in the
statistics collected for decreasing experimental resolution in two regimes: Near peaks
(figure~\ref{fig2}.(a)-(c)) and far apart peaks (figure~\ref{fig2}.(d)-(f)).
It is remarkable that a resolution of the order of magnitude of $b$, which is expected to assume rather small values for classical oscillators, suffices to hide the quantum interference even in the regime of near peaks (first line of figure~\ref{fig2}) when the interference effects are accentuated. According to these results, low resolutions guarantee the emergence of the quasi-determinism -- associated with the delta-like probability distributions -- even for quantum states with no classical analogue, such as the Schrödinger cat. More importantly, this explains the quantum-to-classical transition for closed few degrees of freedom systems.

Of course, for those who interpret the wave function as describing physical reality, the particle delocalization problem would keep existing (before the measurement is performed) even when we are not allowed to experimentally observe it. In this case, decoherence would be mandatorily invoked to destroy quantum coherences and hence settle the problem. However, one must realize that decoherence is proved to be associated with an exponential (in some cases Gaussian) death of quantum coherences, not with the {\em exactly disappearance} of them. Then, from a formal point of view delocalization is always there and the interpretational difficulties remain.

\subsection{Low-resolution limit of entanglement}
The next natural challenge for our approach is trying to explain the classical limit of entanglement. In
the beginning of this work we have given an example in which entanglement does not vanish as $\hbar\to 0$. This has been shown to occur only in a trivial regime of very short times. Would this limit emerge from low-resolution observations? Now we show that the answer for this question appears to be affirmative.

Let us consider the Schmidt decomposition $\sum_i\sqrt{p_i}|\Phi_i\rangle\otimes |\Theta_i\rangle$
for a pure bipartite system $|\psi\rangle$. The respective density operator reads
\begin{eqnarray}\label{psiSchmidt}
|\psi\rangle\langle
\psi|=\sum\limits_{i,j}\sqrt{p_ip_j}\,|\Phi_i\rangle\langle
\Phi_j|\otimes|\Theta_i\rangle\langle\Theta_j|,
\end{eqnarray}
with $\sum_ip_i=1$, for the joint pure state of two subsystems $\Phi$ and $\Theta$. Since
$\{|\Phi_i\rangle\otimes|\Theta_i\rangle\}$ denotes a bi-orthonormal basis, we have $\langle
\Phi_i|\Phi_j\rangle=\delta_{ij}$ and $\langle \Theta_i|\Theta_j\rangle=\delta_{ij}$, which in the position representation are written as
\begin{eqnarray}
\sum\limits_{k}\delta x \,\,\Phi_i^*(x_k)\Phi_j(x_k)=\delta_{ij}, \\
\sum\limits_{k}\delta x
\,\,\Theta_i^*(x_k)\Theta_j(x_k)=\delta_{ij}.
\end{eqnarray}
In the strict classical limit -- defined by low resolutions and well localized wave functions
($\delta x\gg \Delta x$) -- the sums above possess only one term, that is,
\begin{eqnarray}\label{PsiThetaCL}
\Phi_i^*(x_k)\Phi_j(x_k)=\frac{\delta_{ij}}{\delta x}, \\
\Theta_i^*(x_k)\Theta_j(x_k)=\frac{\delta_{ij}}{\delta x}.
\end{eqnarray}
This assumption can be verified straightforwardly, e.g., for coherent and Fock bases, for which wave functions presents Gaussian modulation factors such as $e^{-x^2/b^2}$, where $b=\sqrt{\hbar/m\omega}$. In fact, for $\delta x \sim 10\,b$ it can be shown that (\ref{PsiThetaCL}) is exact.
The probability distribution for (\ref{psiSchmidt}) is then given by
\begin{eqnarray}
|\langle x_k x_{k'}|\psi\rangle|^2\delta
x^2=\sum\limits_{i,j}&\sqrt{p_ip_j}&\Big[\Phi_i^*(x_k)\Phi_j(x_k)\,\delta x\Big]\Big[\Theta_i^*(x_{k'})\Theta_j(x_{k'})\,\delta
x\Big]. \qquad
\end{eqnarray}
By (\ref{PsiThetaCL}) we see that only diagonal terms survive in this expression. The result is a
probability distribution which can be show to derive from the state
\begin{eqnarray}
|\psi\rangle\langle \psi|=\sum\limits_{i}p_i\,|\Phi_i\rangle\langle
\Phi_i|\otimes|\Theta_i\rangle\langle\Theta_i|,
\end{eqnarray}
which properly defines a fully separable quantum state.

Note, at last, that our results derive from assumptions related to both the wave function localization for short times and the experimental resolution. No hypothesis on $\hbar/S$ has been required at all, this being an important difference from most approaches.

\section{Summary and concluding remarks}

In this paper the quantum-to-classical transition has been addressed from an observational viewpoint. By  analytically solving a unitary quantum dynamics rich in phenomena belonging exclusively to the quantum realm such as coherent superpositions and entanglement it has been shown that classical trajectories can emerge from the quantum formalism only at an approximative level. We have seen that the limit $\hbar\to 0$ is not able to make either quantum expectations precisely reduce to Newtonian trajectories or the entanglement vanishes. The only way of succeeding in this task is to require some margin of error within which classical and quantum results can reliably be assumed to coincide. In this case, Newtonian mechanics and hence the concept of determinism is expected to apply for the motion description, so that the notion of a quasi-deterministic behavior as defined in~\cite{angelo07} turns out to be rather appropriate. In addition, note that in this context quantum mechanics is argued to be {\em the universal} theory, since it not only correctly describes the microscopic world but also reproduces -- although approximately -- the classical results for some regime of parameters.

In order to obtain deep insights concerning the relevance of the observational aspects of the quantum-to-classical transition we have introduced the experimental resolution in the quantum formalism by discretizing the structure associated with the probability distributions in the position representation. As a result we have demonstrated generically that in the low-resolution limit i-) Newton's second law arises from the Heiserberg equations of motion, ii-) quantum superpositions are shown not to be experimentally discernible from localized wave functions supporting the concept of quasi-determinism, and iii-) quantum entangled distributions experimentally coalesce to those corresponding to fully separable states.

As a final remark it is important to declare that while our analysis does not dismiss the demonstrated relevance of the EID program in approaching several aspects of the quantum-to-classical transition it strongly emphasizes, together with other contributions~\cite{kofler07,angelo07}, the importance of observational aspects for the subject. Yet, the best of our knowledge, this is the first time that such a general demonstration is given which is able to explains some puzzling problems of the quantum-to-classical transition without appealing either for the EID or for reformulations on the Schrödinger equation. We then conclude that classical physics, which here is claimed to be an approximative description of nature, well succeeds in explain macroscopic motion for several reasons, among which we have to include the low-resolution power of our spectacles.

\section*{Acnowledgements}
The author acknowledges A. D. Ribeiro for carefully reading the initial version of this manuscript and for interesting discussions. The financial support provided by Conselho Nacional de Desenvolvimento Científico e Tecnológico (CNPq) under process 471072/2007-9 is also gratefully acknowledged.


\end{document}